\documentclass[aps,pra,showpacs,twocolumn]{revtex4}
\usepackage{graphicx}
\usepackage{amsthm,amsfonts}
\usepackage{amssymb,amsmath}
\usepackage[applemac]{inputenc}

\begin{document}

%%%%%%%%%%%%%%%%%%%%%%%%%%%%%%%%%%%%%%%%%%%%%%%%%%%%%%%%%%%%%%%%%%%%%
%%%%%%%%%%%%%%%%%%%%%         Title      %%%%%%%%%%%%%%%%%%%%%%%%%%%
%%%%%%%%%%%%%%%%%%%%%%%%%%%%%%%%%%%%%%%%%%%%%%%%%%%%%%%%%%%%%%%%%%%%%

\title{Sending classical information through relativistic quantum channels}

%%%%%%%%%%%%%%%%%%%%%%%%%%%%%%%%%%%%%%%%%%%%%%%%%%%%%%%%%%%%%%%%%%%%%
%%%%%%%%%%%%%%%%%%%%     Authors & Addresses  %%%%%%%%%%%%%%%%%%%%%%%
%%%%%%%%%%%%%%%%%%%%%%%%%%%%%%%%%%%%%%%%%%%%%%%%%%%%%%%%%%%%%%%%%%%%%

\author{Andr\'e G. S. Landulfo}
\email{andre.landulfo@ufabc.edu.br}
\affiliation{Centro de Ci\^{e}ncias Naturais e Humanas, Universidade Federal do ABC, Rua
Santa Ad\'{e}lia 166, 09210-170, Santo Andr\'{e}, S\~{a}o Paulo, Brazil}
\author{Adriano C. Torres}
\email{actorres@ift.unesp.br}
\affiliation{Instituto de F\'\i sica Te\'orica, Universidade Estadual Paulista,
Rua Dr. Bento Teobaldo Ferraz, 271 - Bl. II, 01140-070, S\~ao Paulo, SP, Brazil }

\date{\today}

%%%%%%%%%%%%%%%%%%%%%%%%%%%%%%%%%%%%%%%%%%%%%%%%%%%%%%%%%%%%%%%%%%%%%%
%%%%%%%%%%%%%%%%%%%           Abstract            %%%%%%%%%%%%%%%%%%%%
%%%%%%%%%%%%%%%%%%%%%%%%%%%%%%%%%%%%%%%%%%%%%%%%%%%%%%%%%%%%%%%%%%%%%%

\begin{abstract}
We investigate how special relativity influences the transmission of classical information through quantum channels by evaluating the Holevo bound when the sender and the receiver are in (relativistic) relative motion. By using the spin degrees of freedom of spin-$1/2$ fermions to encode the classical information we show that, for some configurations, the accessible information in the receiver can be increased when the spin detector moves fast enough. This is possible by allowing the momentum wave packet of one of the particles to be wide enough while the momentum wave packets of other particles are kept relatively narrow. In this way, one can take advantage of the fact that boosts entangle the spin and momentum degrees of freedom of spin-$1/2$ fermions to increase the accessible information in the former. We close the paper with a discussion of how this relativistic quantum channel cannot in general be described by completely positive quantum maps.
\end{abstract}


\maketitle

%%%%%%%%%%%%%%%%%%%%%%%%%%%%%%%%%%%%%%%%%%%%%%%%%%%%%%%%%%%%%%%%%%%%%%%
%%%%%%%%%%%%%%%%%%%%%         Text Body          %%%%%%%%%%%%%%%%%%%%%%
%%%%%%%%%%%%%%%%%%%%%%%%%%%%%%%%%%%%%%%%%%%%%%%%%%%%%%%%%%%%%%%%%%%%%%%

\section{Introduction}

Remarkable effects are found in information processing when communication channels are allowed to be quantum mechanical. Examples of these are fast quantum algorithms~\cite{fastquantalg}, quantum  teleportation~\cite{tele1}, quantum cryptography~\cite{cripto1, cripto2, cripto3}, dense coding~\cite{densecoding1} and quantum error correction~\cite{errorcorrection}. Quantum information theory~\cite{nielsen&chuang} commonly deals only with nonrelativistic systems. However, a relativistic treatment is relevant not only to the logical completeness of the theory but also to the disclosure of new physical effects and bounds that arise in information transfer and processing when there is relative motion between the parts that trade information~\cite{Peres&Terno2003,Peres&Terno2004}. Moreover, a better understanding of the relativistic extension of quantum information theory may shed light on several important conceptual issues, for instance, the black hole information ``paradox"~\cite{Hawking1976,Wald2001}. For the aforementioned reasons, a great deal of attention has been paid to quantum information theory in the context of special relativity \cite{PST02, GA02}, the Unruh effect \cite{Fetal, FKMB11, LM09b, CLMS10}, and black holes \cite{PJ08,HBK12,ML10}. Recently, an experimental setup in which a relativistic formulation of quantum information theory may be important was proposed. It consists in using free-space transmission of photons between ground stations and satellites in order to test quantum mechanics for large space distances and, eventually, to implement quantum information protocols in global  scales~\cite{satellite1, satellite2, satellite3, satellite4}. In previous works, we have studied how special relativity affects the correlations between an entangled pair of both fermions~\cite{LM2009} and photons~\cite{LMT2010} by examining the Clauser-Horne-Shimony-Holt Bell inequality~\cite{CHSH1969} when the detectors are moving. Here, we adopt a rather information-theoretic approach and analyze the classical capacity of a relativistic quantum channel in comparison to that of a nonrelativistic one.

One of the keystones of quantum information is the indistinguishability of arbitrary quantum states: given two non-orthogonal quantum states, one cannot distinguish between them with full reliability by making any measurement, a result that is easily shown to be equivalent to the no-cloning  theorem~\cite{nocloning1,nocloning2,nielsen&chuang}. More precisely, let $X=1,...,n$ be an index that indicates each of the states from the set $\left\{ \rho_1, \ldots , \rho_n \right\}$, and suppose that these states are prepared according to a probability distribution $p_1, \ldots, p_n$. An experimentalist performs a measurement described by the positive operator-valued measure (POVM) $\left\{ E_1, \ldots, E_m \right\}$ and is supposed to infer the state $X$ that was prepared from the measurement outcome $Y=1, \ldots, m$. A good measure of how much information the experimentalist can gain about the state through this procedure is given by the mutual information $I \left( X:Y \right)$, which can be defined as
\begin{equation}
 I \left( X:Y \right) = H(X) + H(Y) - H \left( X,Y \right).
\end{equation}
Here, $H(X)$ and $H(Y)$ are the Shannon entropies associated with the probability distribution $p_1, \ldots, p_n$ of the preparation procedure and the probability distribution $\{{\rm tr}(E_1\rho),..., {\rm tr}(E_n\rho)\}$ of the measurement outcomes,  where $\rho\equiv \sum_{x=1}^{n}p_x \rho_x$,  respectively, and $H \left( X,Y \right)$ is their joint entropy. It is well known that $I \left( X:Y \right) \leq H(X)$ and that one can infer $X$ from $Y$ if and only if $I \left( X:Y \right) = H(X)$. The closer $I \left( X:Y \right)$ gets to $H(X)$, the more accurately it is possible to infer $X$ from $Y$. Of course the mutual information depends on what measurement the experimentalist decides to perform, that is, on the POVM chosen. To avoid this  indeterminacy, we define the accessible information to be the maximum of the mutual information over all possible measurement schemes. Although no general method for calculating the accessible  information is known, it is possible to prove a very important upper bound known as the Holevo bound~\cite{Holevo1973}. It states that, for any measurement the experimentalist may do, the inequality below holds:
\begin{equation}
I \left( X:Y \right) \leq \chi(\rho)\equiv S\left( \sum_{x=1}^{n}p_x \rho_x \right) - \sum_{x=1}^{n} p_x S\left( \rho_x \right),
\label{holevobound}
\end{equation}
where $S \left( \omega \right) = - \textup{tr} \left( \omega \, \log_2 \omega \right)$ is the von Neumann entropy of the quantum state $\omega$. It is easy to show~\cite{nielsen&chuang} that $\chi(\rho) \leq H(X)$ and, therefore, inequality~(\ref{holevobound}) implies that one  qubit contains at most one bit of information. The Holevo bound is especially relevant due to the Holevo-Schumacher-Westmoreland (HSW) theorem, independently proved by Schumacher and Westmoreland~\cite{Schumacher&Westmorland1997} and by Holevo~\cite{Holevo1998}, according  to which the rate $\chi(\rho)$ is asymptotically achievable and thus can be used to obtain an expression for the classical (product state) capacity of a quantum channel. Hence, in order to study the transmission of classical information through relativistic quantum channels, it is interesting to analyze the Holevo bound in a typical relativistic quantum communication setup. Here this is done by setting the detector in relative motion with respect to the preparation apparatus.

The indistinguishability of non-orthogonal quantum states mentioned above together with fact that any attempt to distinguish them ends up only imparting a disturbance to the states is what motivates their use in cryptographic protocols. This however is not the only setup in which non-orthogonal states play a relevant role. As it was first shown in~\cite{fucs97} there are some noisy quantum channels in which the channel capacity is only achieved by the use of non-orthogonal states. In the present paper we will show that when the parts that trade information are in relative motion (and there is not any external noise afflicting the states), there is a certain class of non-orthogonal states for which the Holevo bound increases when compared to its value when the sender and the receiver are at rest relative to each other. This suggests that the relative motion may actually help to increase the capacity of some noisy quantum channels. 

The paper is organized as follows. In Sec.~\ref{nonrelativistic} we analyze the transmission of two and four classical bits through nonrelativistic quantum communication channels. In Sec.~\ref{relativistic} we study how relativity influences the previous quantum communication process. This is done by setting the receiver, Bob, in relative motion with respect to the sender, Alice. In this context  we analyze how to optimize the accessible information on the receiver. In Sec.~\ref{quantummap} we define quantum maps that describe the quantum channels analyzed in the previous sections and show that there are cases in which they fail to be completely positive. Sec.~\ref{finalremarks} is dedicated to our final remarks. We adopt natural units $c=\hbar=1$ unless stated otherwise.

\section{Nonrelativistic quantum communication setup}
\label{nonrelativistic}
\subsection{Two classical bits}
\label{twobits}
Let us assume that Alice has a classical information source that produces symbols $X=0,1$ according to the  probability distribution $p_0=\lambda, p_1=1-\lambda$, $0\leq \lambda \leq 1$. Depending on the value she obtains for $X$, she prepares a pure quantum state $\psi_X$, chosen from a fixed set $\left\{ \psi_0, \psi_1 \right\}$, of a spin-$1/2$ particle with mass $m$ and then sends it to Bob. He then makes a spin measurement of his choice on that state and has to identify $X$ based on the outcome $Y$. We assume that (see, e.g., Ref.~\cite{Bogolubov1975} for the two-spinor notation used below)
\begin{eqnarray}
\psi_0 ({\bf p}) &=&  \left(\begin{array}{c} f_{\mathbf{k}_0}^{w_0} \left( \mathbf{p} \right) \\ 0 \end{array}\right),
\label{psi1}
\\
\psi_1({\bf p}) &=& \cos \theta
 \left(
 \begin{array}{c} 
f_{\mathbf{k}_1}^{w_1}(\mathbf{p})\\
0 \\  
\end{array} \right)
  + \sin \theta
  \left(
 \begin{array}{c} 
0\\
f_{\mathbf{k}_1}^{w_1}(\mathbf{p})\\  
\end{array} \right),
\label{psi2}
\end{eqnarray}
where
\begin{equation}
f_{\mathbf{k}_i}^{w_i} \left( \mathbf{p} \right) = \pi^{-\frac{3}{4}} w_i^{-\frac{3}{2}} \, \exp \left[ -\left( \mathbf{p} - \mathbf{k}_i \right)^2/2w_i^2 \right].
\label{pacotegaussiano}
\end{equation}
Here, the parameters $w_i\in \mathbb{R}_+$ and $\mathbf{k}_i = \left( k_i,0,0 \right)$, $i=0,1$, give the particle's momentum dispersion and the average momentum, respectively. Note that $\psi_1$ can be written as $$\psi_1({\bf p})=f_{\mathbf{k}_1}^{w_1}(\mathbf{p}) \left(
 \begin{array}{c} 
\cos \theta\\
\sin \theta\\  
\end{array} \right),$$ 
making manifest its direct product structure.

\begin{figure}[t]
\begin{center}
\includegraphics[height=0.25\textheight]{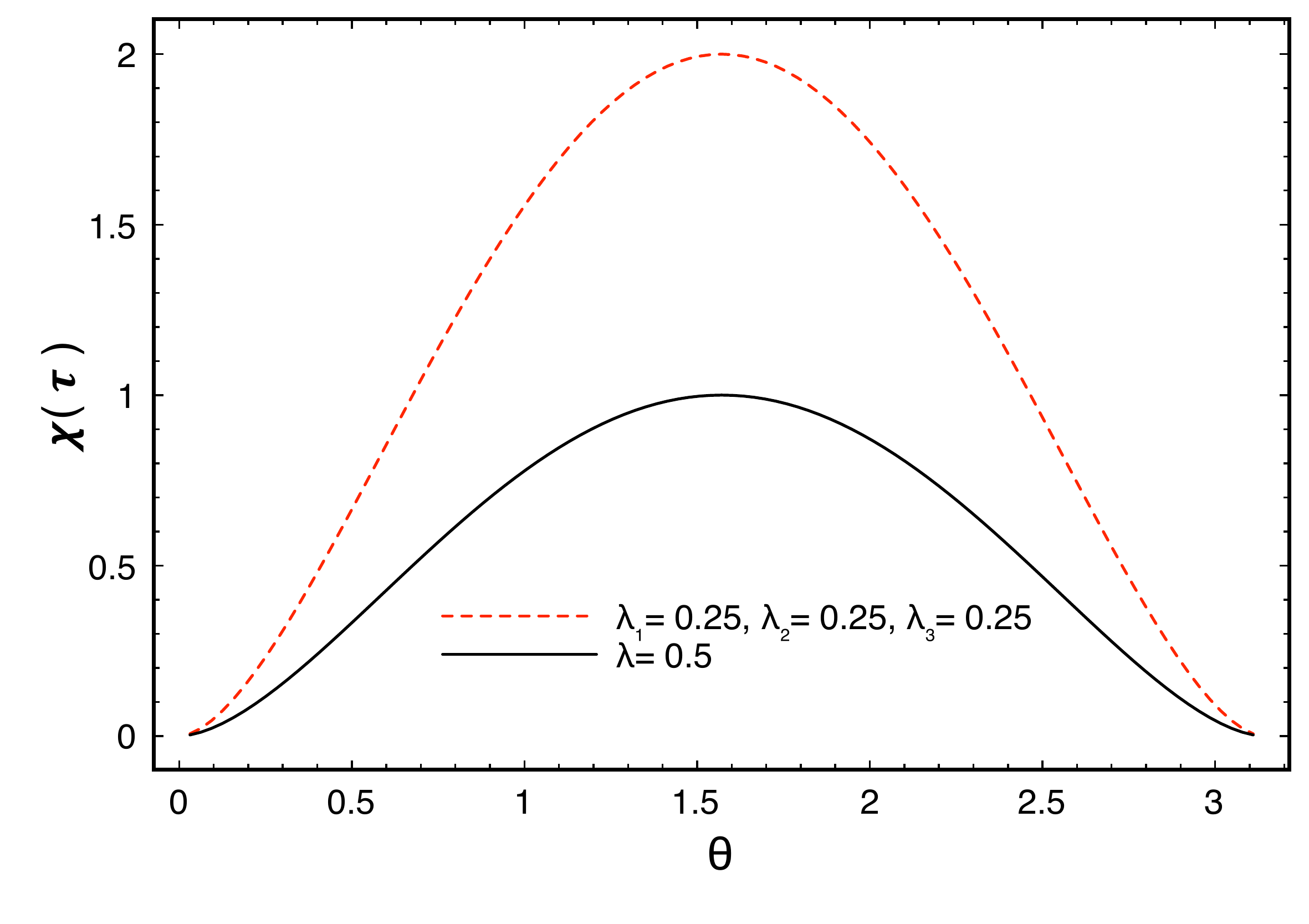}
\caption{(Color online) The graph shows the behavior of the Holevo bound when Bob is at rest with respect to Alice as a function of the angle $\theta$, which characterize the spin part of the second state, for both the transmission of two (full line) and four (dashed line) classical bits. }
\label{plot0}
\end{center}
\end{figure}

First we summarize what happens when Bob is at rest relative to Alice, which is the typical quantum information scenario. All spin measurement results can be predicted through the reduced spin density operator, obtained by tracing out the momenta,
\begin{equation}
\tau \equiv \int \mbox{d}{\bf p} \; \rho \left( \mathbf{p}, \mathbf{p} \right),
\end{equation}
where
\begin{equation}
\rho \left( \mathbf{p}, \mathbf{\tilde{p}} \right) = \lambda \, \psi_0 (\mathbf{p}) \psi_0(\mathbf{\tilde{p}})^\dag + (1-\lambda) \, \psi_1 (\mathbf{p}) \psi_1 (\mathbf{\tilde{p}})^\dag
\label{completerho}
\end{equation}
is the complete density operator of the system. We thus obtain
\begin{equation}
\tau = \lambda   \left(
 \begin{array}{cc} 
1& 0\\
0& 0\\  
\end{array} \right)+ (1-\lambda)  \left(
 \begin{array}{cc} 
 \cos^2 \theta & \cos \theta \sin \theta \\
\cos \theta \sin \theta & \sin^2 \theta\\ 
\end{array} \right).
\label{reducedspin}
\end{equation} 
Note that, due to the linearity of the trace, we can write $\tau \equiv \lambda \, \tau_0 + (1-\lambda) \, \tau_1$, where
\begin{equation}
\tau_0 \equiv \int \mbox{d}{\bf p} \; \psi_0(\mathbf{p}) \psi_0 (\mathbf{p})^\dag=   \left(
 \begin{array}{cc} 
1& 0\\
0& 0\\  
\end{array} \right),
\label{tau0}
\end{equation} 
and
\begin{equation}
\tau_1 \equiv \int \mbox{d}{\bf p} \; \psi_1(\mathbf{p}) \psi_1 (\mathbf{p})^\dag=   \left(
 \begin{array}{cc} 
 \cos^2 \theta & \cos \theta \sin \theta \\
\cos \theta \sin \theta & \sin^2 \theta\\ 
\end{array} \right).
\label{tau1}
\end{equation} 
As can be easily seen from Eq.~(\ref{reducedspin}), $\tau$ is just a statistical mixture of the pure spin states 
$\phi_{\uparrow}$ and 
\begin{equation}
\phi_{\theta}\equiv \cos \theta \, \phi_{\uparrow} + \sin \theta \, \phi_{\downarrow},
\label{phitheta}
\end{equation}
where $\phi_{\uparrow}\equiv \left(
 \begin{array}{c} 
1\\
0 \\  
\end{array} \right)$ and $\phi_{\downarrow}\equiv\left(
 \begin{array}{c} 
0\\
1\\  
\end{array} \right)$ are the eigenvectors of $S_{\rm z}$ with eigenvalues $1/2$ and $-1/2$, respectively. Note that  $\phi_{\theta}$ is an eigenstate of ${\bf S}\cdot {\bf n}$ where ${\bf n}=(\sin 2\theta,0,\cos 2\theta)$, ${\bf S}\equiv\boldsymbol{\sigma}/2$, and  $\boldsymbol{\sigma} = \left( \sigma_{\rm x}, \sigma_{\rm y}, \sigma_{\rm z} \right)$ is the Pauli vector. Thus, we can say that the momentum degrees of freedom play absolutely no role in this case. As $\tau_1$ and $\tau_2$ correspond to pure states, we have $S \left( \tau_1 \right) = S \left( \tau_2 \right) = 0$, so that
\begin{equation}
\chi \left( \tau \right) = S \left( \tau \right) = - \sum_{l=\pm} \beta_l \, \log_2 \beta_l,\end{equation}
where $\beta_\pm$ are the eigenvalues of $\tau$, namely,
\begin{equation}
\beta_\pm = \frac{1}{2} \pm \frac{1}{2} \sqrt{1 + 4 \sin^2 \theta \left( \lambda^2-\lambda \right)}.
\end{equation}
Note that $\chi(\tau)$ has periodicity $\pi$ with respect to $\theta$. It is straightforward to verify that the Holevo bound reaches its maximum value when $\theta=\pi/2$, corresponding to orthogonal spin states (see Fig.~\ref{plot0}). At this point, and only at this point, it is possible for Bob to determine with certainty which state Alice has prepared, which he does simply by measuring $S_z$. This is actually a general property of orthogonal states, which can always be completely distinguished by appropriate measurements. By contrast, the minimum is attained at $\theta=0$ and $\theta=\pi$, which means that the states $\psi_0$ and $\psi_1$ are identical (up to a phase) when it comes solely to spin and, therefore, they cannot be distinguished at all. 

\subsection{Four classical bits}
\label{fourbits}
Now, let us see how the above results generalize to the case where more classical bits are sent through the quantum communication channel. For this purpose, let us suppose now that Alice has a classical information source that produces symbols $\tilde{X}=00,01,10,11$ according to the  probability distribution $p_{00}=\lambda_1, p_{01}=\lambda_2, p_{10}=\lambda_3$, $p_{11}=\lambda_4=1-\lambda_1-\lambda_2-\lambda_3$, $0\leq\lambda_{\tilde{l}}\leq 1,$ $\tilde{l}=1,2,3,4$. As in the previous case, Alice can choose pure quantum states from the set $\left\{ \psi_0, \psi_1 \right\}$, where $\psi_0$ and $\psi_1$ are given by Eqs.~(\ref{psi1}) and~(\ref{psi2}), respectively. Therefore, depending on the value of $\tilde{X}$, Alice prepares one of the four product states 

\begin{eqnarray}
\psi_{00}({\bf p},\tilde{{\bf p}} )\!\!&\equiv&\! \!\psi_0({\bf p})\otimes \psi_0(\tilde{{\bf p}}),\;  \psi_{01}({\bf p},\tilde{{\bf p}} )\equiv \psi_0({\bf p})\otimes \psi_1(\tilde{{\bf p}}),  \nonumber \\
&&\nonumber \\
\psi_{10}({\bf p},\tilde{{\bf p}} )\!\!&\equiv&\!\! \psi_1({\bf p})\otimes \psi_0(\tilde{{\bf p}}), \;
 \psi_{11}({\bf p},\tilde{{\bf p}} )\equiv \psi_1({\bf p})\otimes \psi_1(\tilde{{\bf p}}),\nonumber \\
 \end{eqnarray}
and sends it to Bob. He then makes a spin measurement of his choice on that state and has to identify $\tilde{X}$ based on the outcome $\tilde{Y}$. The total density operator of the system is given by
\begin{eqnarray}
&& \tilde{\rho} \left(\mathbf{p}, \mathbf{\tilde{p}},\mathbf{p}', \mathbf{\tilde{p}}' \right) = \lambda_1 \psi_{00}({\bf p},\tilde{{\bf p}} )\psi_{00}^\dag({\bf p}',\tilde{{\bf p}}' ) \nonumber \\ 
&&+ \lambda_2 \, \psi_{01}({\bf p},\tilde{{\bf p}} )\psi_{01}^\dag({\bf p}',\tilde{{\bf p}}' ) 
+ \lambda_3 \, \psi_{10}({\bf p},\tilde{{\bf p}} )\psi_{10}^\dag({\bf p}',\tilde{{\bf p}}' ) \nonumber \\
&&+\lambda_4 \, \psi_{11}({\bf p},\tilde{{\bf p}} )\psi_{11}^\dag({\bf p}',\tilde{{\bf p}}' ).
\label{totdens4}
\end{eqnarray}
If the momentum degrees of freedom are traced out in Eq.~(\ref{totdens4}), we obtain the reduced spin density operator 

\begin{eqnarray}
\tilde{\tau} \equiv \int \mbox{d}{\bf p} \mbox{d}\tilde{{\bf p}} \; \tilde{\rho} \left( \mathbf{p},\tilde{ \mathbf{p}},\mathbf{p}, \tilde{\mathbf{p}}\right),
\end{eqnarray}
which can be written as 
\begin{equation}
\tilde{\tau}= \lambda_1 \tau_0\otimes \tau_0 + \lambda_2 \tau_0\otimes \tau_1 \nonumber \\
+ \lambda_3\tau_1\otimes \tau_0 + \lambda_4 \tau_1\otimes \tau_1,
\end{equation}
and from which all spin measurement results can be predicted. The density operators $\tau_0$ and $\tau_1$ are given in Eqs.~(\ref{tau0}) and~(\ref{tau1}), respectively.  Now, by using that $S(\omega_1\otimes \omega_2)=S(\omega_1)+S(\omega_2)$ for any density matrices $\omega_1$ and $\omega_2$ and that $S(\tau_i)=0,$ we can write $\chi\left(\tilde{\tau}\right)$ as 
\begin{equation}
\chi \left( \tilde{\tau} \right) = S \left(\tilde{ \tau} \right) = - \sum_{\tilde{l}=1}^{4} \tilde{\beta}_{\tilde{l}} \, \log_2 \tilde{\beta}_{\tilde{l}},\end{equation}
where $\tilde{\beta}_{\tilde{l}}$ are the eigenvalues of $\tilde{\tau}.$ In Fig~\ref{plot0}, $\chi \left( \tilde{\tau} \right)$ is plotted when $\lambda_1=\lambda_2=\lambda_3=1/4$. We can see that, as in the case of two bits, its maximum value is attained when $\theta=\pi/2.$ 

\section{Relativistic Quantum Communication Optimization }
\label{relativistic}
Now we turn our attention to the relativistic case, in which Bob moves with three-velocity $\mathbf{v} = \left( v,0,0 \right)$ relative to Alice.
\subsection{Two classical bits}
\label{twobitsrel}
Let us suppose first that Alice wants to transmit two bits of classical information. Thus, as explained in Section~\ref{twobits}, she prepares the state $\rho$ given in Eq.~(\ref{completerho}) and sends it to Bob. Due to his motion, Bob sees the state $\rho$ prepared by Alice unitarily transformed in his proper frame as 
\begin{equation}
\rho' \left( \mathbf{p}, \mathbf{\tilde{p}} \right) = \lambda \, \psi_0' (\mathbf{p}) \psi_0' (\mathbf{\tilde{p}})^\dag + (1-\lambda) \, \psi_1' (\mathbf{p}) \psi_1' (\mathbf{\tilde{p}})^\dag
\label{rho`2}
\end{equation}
where~\cite{Halpern1968, Weinberg1996}
\begin{equation}
\psi_i' \left( \mathbf{p} \right) \equiv \left(U(\Lambda)\psi_i\right)({\bf p})
\end{equation}
with
\begin{eqnarray}
 \left(U(\Lambda)\psi_i\right)({\bf p})\equiv \sqrt{\frac{\left( \Lambda^{-1}p \right)^0}{p^0}} \, D \left( \Lambda, \Lambda^{-1}p \right) \psi_i \left( \Lambda^{-1} \mathbf{p} \right)\!\!,
\label{booststate}
\end{eqnarray}
$p = \left( \sqrt{\mathbf{p}^2+m^2},\mathbf{p} \right)$, and  $\Lambda^{-1} \mathbf{p}$ denoting the spatial part of the four-vector $\Lambda^{-1} p$. The Wigner rotation is given by

\begin{eqnarray}
D \left( \Lambda, q \right) &=& \frac{\cosh(\alpha/2)\,(q^0+ m)\sigma^0}{[(p^0+m)(q^0+m)]^{1/2}}\nonumber \\
&+& \frac{\sinh(\alpha/2)[{\bf q}\cdot {\bf e}\; \sigma^0+i({\bf e}\times {\bf q})\cdot \boldsymbol{\sigma}]}{[(p^0+m)(q^0+m)]^{1/2}},
\label{wignerrotation}
\end{eqnarray}
where $\alpha = - \tanh^{-1} v$, $q\equiv \Lambda^{-1}p,$ $\sigma^0=I$ is the identity matrix, and  $\mathbf{e}$ gives the direction of the boost, which in our case is ${\bf e}_x$, so that
\begin{equation}
\Lambda = \left(\begin{array}{cccc}
\cosh \alpha & \sinh \alpha & 0 & 0 \\ 
\sinh \alpha & \cosh \alpha & 0 & 0 \\ 
0 & 0 & 1 & 0 \\ 
0 & 0 & 0 & 1
\end{array}\right).
\end{equation}
By using Eq.~(\ref{wignerrotation}) in Eq.~(\ref{booststate}) with $\psi_i$ given in Eqs.~(\ref{psi1}) and~(\ref{psi2}), we obtain 
\begin{equation}
\psi_0' (\mathbf{p}) =\left(\begin{array}{c} a_{\mathbf{k}_0}^{w_0} (\mathbf{p}) \\ b_{\mathbf{k}_0}^{w_0} (\mathbf{p}) \end{array}\right)
\label{psi0prime}
\end{equation}
and
\begin{equation}
\psi_1' (\mathbf{p}) = \cos \theta \left(\begin{array}{c} a_{\mathbf{k}_1}^{w_1} (\mathbf{p}) \\ b_{\mathbf{k}_1}^{w_1} (\mathbf{p}) \end{array}\right) + \sin \theta \left(\begin{array}{c} - b_{\mathbf{k}_1}^{w_1} (\mathbf{p}) \\ a_{\mathbf{k}_1}^{w_1} (\mathbf{p})^* \end{array}\right),
\label{psi1prime}
\end{equation}
where the momentum wave packets are given by
\begin{eqnarray*}
a_{\mathbf{k}_i}^{w_i} (\mathbf{p})&=& K \, f_{\mathbf{k}_i}^{w_i} (\mathbf{q}) \left[ C (q^0+m) + S (q_x+iq_y) \right],  \\ 
\\
b_{\mathbf{k}_i}^{w_i} (\mathbf{p})&=& K \, f_{\mathbf{k}_i}^{w_i} (\mathbf{q}) \, S q_z,
\end{eqnarray*}
with 
\begin{eqnarray*}
K &\equiv& (q^0/p^0)^{1/2}/ [(q^0+m)(p^0+m)]^{1/2}, \\
C &\equiv&\cosh \left(\alpha/2 \right), \\
S &\equiv& \sinh \left( \alpha/2 \right).
\end{eqnarray*}
By tracing out the momentum degrees of freedom we obtain the reduced spin operator in Bob's frame
\begin{equation}
\tau' = \lambda \, \tau_0' + (1-\lambda) \tau_1',
\label{tau'v1}
\end{equation}
where
\begin{equation}
\tau' \equiv \int \mbox{d}{\bf p} \; \rho' (\mathbf{p}, \mathbf{p}),
\label{tau`v2}
\end{equation}
\begin{equation}
\tau_i' \equiv \int \mbox{d}{\bf p} \; \psi_i' (\mathbf{p}) \psi_i' (\mathbf{p})^\dag,
\end{equation}
and $ \rho' (\mathbf{p}, \mathbf{p})$ is given in Eq~(\ref{rho`2}). Explicitly, we have
\begin{eqnarray}
\tau_0'&=& \left(\begin{array}{cc} 1-V(\alpha) & 0 \\ 0 & V(\alpha) \end{array}\right),
\label{tau1'}
\\
\tau_1' &=& \left(\begin{array}{cc} A(\alpha) & B(\alpha) \\ B(\alpha) & 1-A(\alpha) \end{array}\right),
\label{tau2'}
\end{eqnarray}
where
\begin{eqnarray}
A(\alpha) &=& \cos^2 \theta \left[1-U(\alpha) \right] + \sin^2 \theta \, U(\alpha),
\label{Aalpha}
\\
B(\alpha) &=& \cos \theta \sin \theta \, \left[1-4 \, U(\alpha) \right]
\label{Balpha}
\end{eqnarray}
with $V(\alpha)$ and $U(\alpha)$ being given by
\begin{equation}
V(\alpha)\equiv\sinh^2 \left(\frac{\alpha}{2} \right) \int\mbox{d}{\bf q} \, \frac{{q_z}^2\left| f_{\mathbf{k}_0}^{w_0} (\mathbf{q}) \right|^2}{(q^0+m)(p^0+m)},
\label{V}
\end{equation}
and 
\begin{equation}
U(\alpha) \equiv \sinh^2 \left( \frac{\alpha}{2} \right) \int \mbox{d} {\bf q} \, \frac{{q_z}^2\left| f_{\mathbf{k}_1}^{w_1} (\mathbf{q}) \right|^2}{(q^0+m)(p^0+m)},
\label{U}
\end{equation}
respectively. In the above equations, we have used the fact that $\mbox{d}{\bf p} / p^0$ is a relativistic invariant and performed the change of variables $q=\Lambda^{-1}p$. Using Eqs.~(\ref{tau1'}) and~(\ref{tau2'}) in Eq~(\ref{tau'v1}),  we can cast $\tau'$ as
\begin{equation}
\tau' \!=\!\left(\!\!\!\begin{array}{cc} \lambda \left(1-V \right) + (1-\lambda) A & (1-\lambda) B \\ (1-\lambda) B& \lambda \, V+ (1-\lambda) (1-A) \end{array}\!\!\!\right)\!\!.
\label{tau'}
\end{equation}
Note that, contrary to $\tau_0$ and $\tau_1$, $\tau_0'$ and $\tau_1'$ are not pure states, so that $S(\tau_0')$ and $S(\tau_1')$ are both non-zero. The Holevo bound in Bob's frame is given by
\begin{equation}
\chi (\tau') = S (\tau') - \lambda \, S (\tau_0') - (1-\lambda) \, S (\tau_1').
\label{hol'}
\end{equation}
By using Eqs.~(\ref{tau1'}), (\ref{tau2'}) and (\ref{tau'}), the above equation can be rewritten as	
\begin{eqnarray}
\chi(\tau') &=& -\sum_{l=\pm} \gamma_l \, \log_{2} \gamma_l + \lambda \sum_{l=\pm} \delta_l\, \log_{2} \delta_l   \nonumber \\
&+& (1-\lambda) \sum_{l=\pm} \epsilon_l \, \log_{2} \epsilon_l,
\label{chi'}
\end{eqnarray}
where, for $l=\pm$, $\gamma_l$, $\delta_l$, and $\epsilon_l$ are the eigenvalues of $\tau'$, $\tau_0'$, and $\tau_1'$, respectively. These can be easily calculated as functions of the integrals $V(\alpha)$ and $U(\alpha)$. In this paper, we proceed a numerical analysis of Eq.~(\ref{chi'}) to examine various aspects of the transmission of classical information through relativistic quantum channels. In order to do that, we rewrite Eqs.~(\ref{V}) and~(\ref{U}) as  

\begin{equation}
V(\alpha) = \frac{\sinh^2 \left( \alpha/2 \right)}{\sqrt{\pi} \; {W_0}^3} \int_{-\infty}^{\infty} \! \! \! \mbox{d} Q_x \int_{0}^{\infty} \! \! \! \mbox{d} Q_r \; G_0 \left( Q_x,Q_r \right),
\label{Valpha}
\end{equation}
and
\begin{equation}
U(\alpha) = \frac{\sinh^2 \left( \alpha/2 \right)}{\sqrt{\pi} \; {W_1}^3} \int_{-\infty}^{\infty} \! \! \! \mbox{d} Q_x \int_{0}^{\infty} \! \! \! \mbox{d} Q_r \; G_1 \left( Q_x,Q_r \right),
\end{equation}
respectively, where we have used Eq.~(\ref{pacotegaussiano}) and cylindrical coordinates with $q_x$ as the symmetry axis, and introduced
\begin{equation}
G_i \left( Q_x, Q_r \right) = \frac{{Q_r}^3 \exp \left\{ - \left[ \left( Q_x - K_i \right)^2 + {Q_r}^2 \right]/{W_i}^2 \right\}}{\left( Q^0+1 \right) \left( Q^0 \cosh \alpha - Q_x \sinh \alpha +1 \right)}.
\end{equation}
In addition, we have defined the normalized non-dimensional variables $Q_r\equiv q_r/m$, $Q_x\equiv q_x/m$, $Q^0=\sqrt{{Q_x}^2+{Q_r}^2+1}$, $W_i\equiv w_i/m$ and $K_i\equiv k_i/m$. 
\begin{figure}[t]
\begin{center}
\includegraphics[height=0.25\textheight]{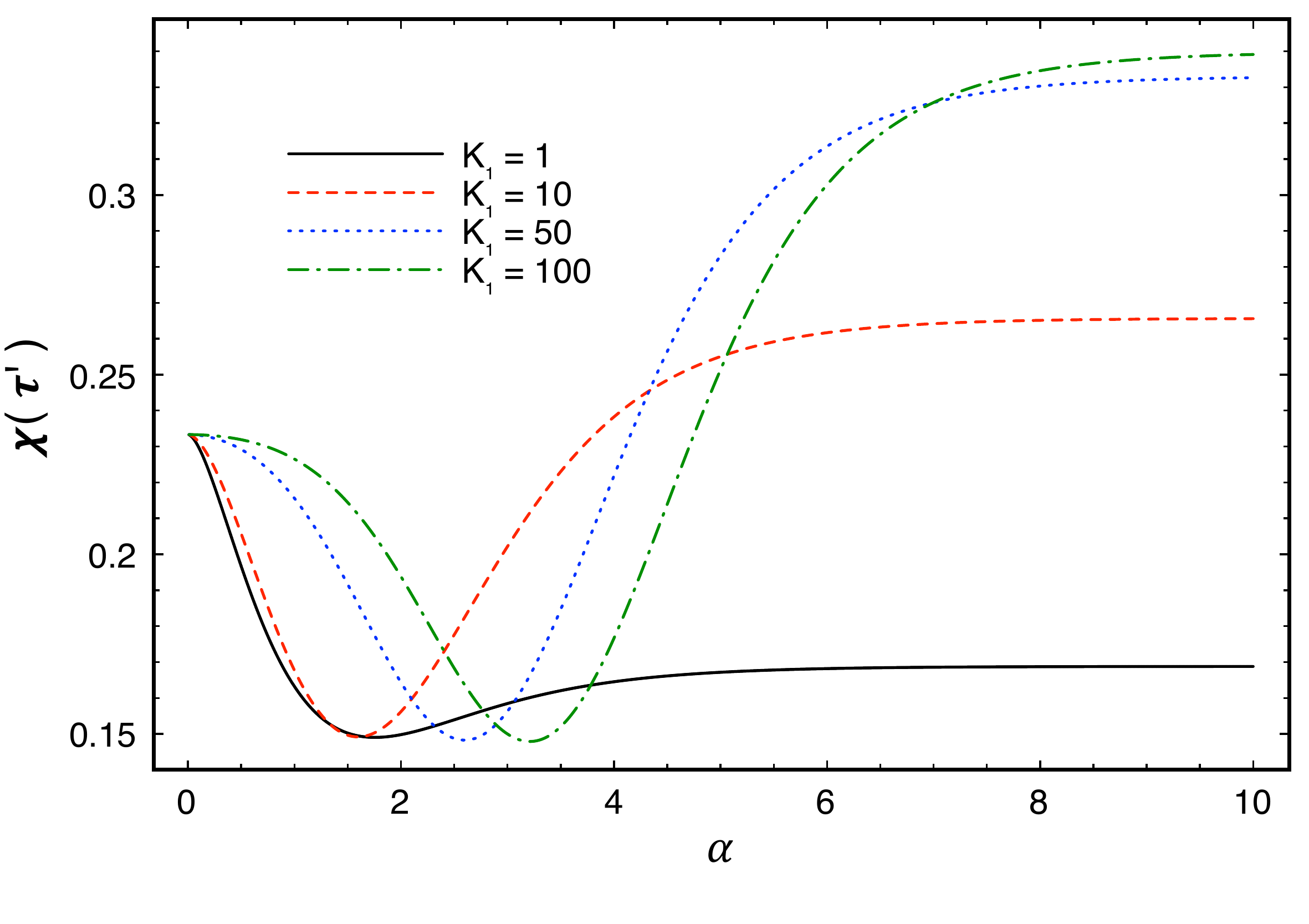}
\caption{(Color online) The Holevo bound $\chi(\tau')$ is plotted as a function of $\alpha=-\tanh^{-1}v$ for different values of the normalized momentum $K_1=k_1/m$. We have fixed $W_0=0.05,$ $W_1=6,$ $\lambda=1/2,$ $K_0=1$, and $\theta=\pi/8$. We can see that although $\chi(\tau')$ decreases with $\alpha$ initially, it begins to increase for larger values of the rapidity, overcoming the value for $\alpha=0$ when $K_1$ is large enough.}
\label{plot1}
\end{center}
\end{figure}

Let us first analyze the behavior of Eq~(\ref{hol'}) [or, equivalently, Eq~(\ref{chi'})]  as a function of the rapidity $\alpha=-\tanh^{-1}v,$ where we recall that $v$ is the relative velocity between Bob and  Alice.  First, it should be noted that if both $W_0$ and $W_1$ were much smaller than the unity, the effects of Bob's velocity would be negligible and, for any value of $\alpha$, $\chi(\tau')$ would have, approximately, its value for $\alpha=0$ (since in this case, $\psi_0$ and $\psi_1$ would be almost momentum eigenstates). However, if we allow $W_1$ to be much larger than $W_0,$ the situation changes considerably, as can be seen in Fig.~\ref{plot1}. There, $\chi(\tau')$ is plotted as a function of the rapidity $\alpha$, for different values of the normalized mean momentum $K_1$ of $\psi_1$, when $W_0=0.05$, $W_1=6$, $\lambda=1/2$, $K_0=1$, and $\theta=\pi/8$. We can see that although $\chi(\tau')$  initially decreases with $\alpha$, it begins to increase for larger values of the rapidity, eventually overcoming the value for $\alpha=0$ when $K_1$ is large enough. Thus, for some codifications, i.e., for some values of $\theta$, the (relativistic) velocity of the receiver can actually {\em increase} his accessible information. This is possible because, when $K_1$ is large, even though $W_1/W_0\approx 10^2$, the states $\psi_0$ and $\psi_1$ are almost orthogonal (despite of the fact that their spin parts, $\phi_{\uparrow}$ and $\phi_{\theta}$, are not), as can be easily checked. When Bob is moving, the momentum and spin degrees of freedom of the original state prepared by Alice are mixed up by the Wigner rotation. As a consequence, Bob can obtain some extra information in the spin degrees of freedom due to the fact that the states $\psi_0$ and $\psi_1$ prepared by Alice are more clearly distinguishable in momentum than in spin.
\begin{figure}[t]
\begin{center}
\includegraphics[height=0.25\textheight]{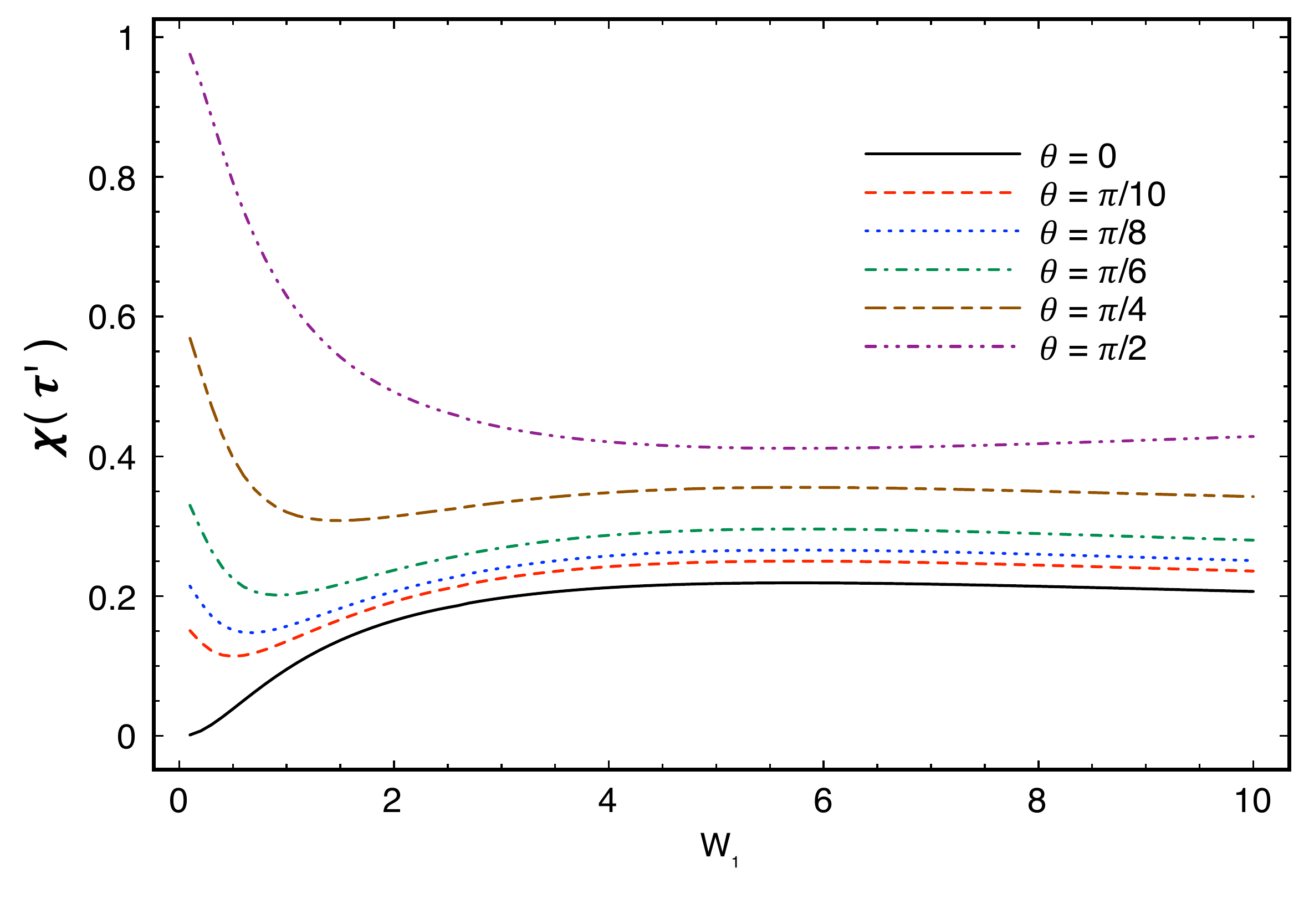}
\caption{\footnotesize{(Color online) The graph shows the Holevo bound $\chi(\tau')$ as a function of $W_1$ for different values of the angle $\theta$ when $\alpha\rightarrow \infty$. We have fixed $K_0=1$, $K_1=10$, $W_0=0.05$, and $\lambda=1/2.$ When $\theta$ is small and $W_1$ is large enough, the accessible information always increases when Bob moves fast enough with respect to Alice. When $\theta$ is larger (in particular, when it is closer to $\pi/2$), the increase in $W_1$ always leads to a decrease in $\chi(\tau')$ and therefore in the accessible information.}} 
\label{plot2}
\end{center}
\end{figure}

In view of the above results, let us study now how the width of the wave packet of the second state influences the accessible information in different codification schemes.  In Fig.~\ref{plot2}, $\chi(\tau')$ is plotted as a function of $W_1$, in the limit $\alpha\rightarrow \infty$, for different values of $\theta$. We have fixed $W_0=0.05,$ $K_0=1,$ $K_1=10$, and $\lambda=1/2$. We can see that when $\theta$ is small and $W_1$ is large enough, the accessible information always increase when Bob moves with relativistic velocities with respect to Alice. When $\theta$ is larger (in particular, when it is close to $\pi/2$), the increase in $W_1$ always leads to a decrease in $\chi(\tau')$ and therefore in the accessible information. This is so because the spin parts, $\phi_{\uparrow}$ and  $\phi_{\theta},$ of the states $\psi_0$ and $\psi_1$ prepared by Alice are already quite distinguishable and thus the entanglement between the spin and momentum degrees of freedom caused by Bob's movement ends up only making the spin degrees of freedom less distinguishable.  It is interesting to note that Alice can codify the information in states $\psi_0$ and $\psi_1$ using $\theta=0$. This makes the states  completely indistinguishable in spin in her frame. Yet, Bob can have a non zero value of $\chi(\tau')$, as can be seen from Fig.~\ref{plot2}. 
Thus, Alice can use the fact that spin and momentum are mixed up in Bob's frame to ``hide" the information in the momentum degrees of freedom in her frame, and let Bob movement make this information available to him in spin. This may be useful when considering noisy quantum channels in the spin degrees of freedom.  
\subsection{Four Classical Bits}
We will analyze now if the above results still apply when Alice sends more classical bits through the quantum channel. For the sake of simplicity, we will describe here what happens when she tries to transmit four bits of classical information to Bob. Thus, as explained in Section~\ref{fourbits}, Alice prepares the state $\tilde{\rho}$ given in Eq.~(\ref{totdens4}) and sends it to Bob. As Bob is moving with respect to her, he sees the state $\tilde{\rho}$ prepared by Alice unitarily transformed in his proper frame as 
\begin{eqnarray}
&& \tilde{\rho}' \left( \mathbf{p}, \mathbf{\tilde{p}},\mathbf{p}', \mathbf{\tilde{p}}' \right) = \lambda_1 \,\psi_{00}'({\bf p},\tilde{{\bf p}} ){\psi'}_{00}^{\dag}({\bf p}',\tilde{{\bf p}}' ) \nonumber \\
&&+ \lambda_2 \, \psi'_{01}({\bf p},\tilde{{\bf p}} ){\psi'}_{01}^{\dag}({\bf p}',\tilde{{\bf p}}' ) 
+ \lambda_3 \, \psi'_{10}({\bf p},\tilde{{\bf p}} ){\psi'}_{10}^{\dag}({\bf p}',\tilde{{\bf p}}' ) \nonumber \\
&&+ \lambda_4 \, \psi'_{11}({\bf p},\tilde{{\bf p}} ){\psi'}_{11}^{\dag}({\bf p}',\tilde{{\bf p}}' ),
\label{totdens4prime}
\end{eqnarray}
where $\psi'_{i j}({\bf p}, \tilde{{\bf p}})\equiv \psi'_i({\bf p})\otimes \psi'_j(\tilde{{\bf p}})$, $i,j=0,1$, and $\psi'_0$ and $\psi'_1$ are given in Eqs.~(\ref{psi0prime}) and~(\ref{psi1prime}), respectively. If we trace out the momentum degrees of freedom in $\tilde{\rho}' \left( \mathbf{p}, \mathbf{\tilde{p}},\mathbf{p}', \mathbf{\tilde{p}}' \right) $ we obtain the density operator
\begin{eqnarray}
\tilde{\tau}' &=& \int \mbox{d}{\bf p} \mbox{d}\tilde{{\bf p}} \; \tilde{\rho}' \left( \mathbf{p}, \mathbf{\tilde{p}},\mathbf{p}, \mathbf{\tilde{p}} \right)
\label{tildetau'v2}
\end{eqnarray}
which can be written as 
\begin{eqnarray}
\tilde{\tau}' &=&\lambda_1 \tau'_0\otimes \tau'_0 + \lambda_2 \tau'_0\otimes \tau'_1 \nonumber \\
&+& \lambda_3\tau'_1\otimes \tau'_0 + \lambda_4 \tau'_1\otimes \tau'_1,
\label{tildetau'2}
\end{eqnarray}
and from which all spin measurement results can be predicted. We recall that $\tau'_0$ and $\tau'_1$ are given in Eqs.~(\ref{tau1'}) and~(\ref{tau2'}), respectively. 

\begin{figure}[t]
\centering
\includegraphics[height=0.25\textheight]{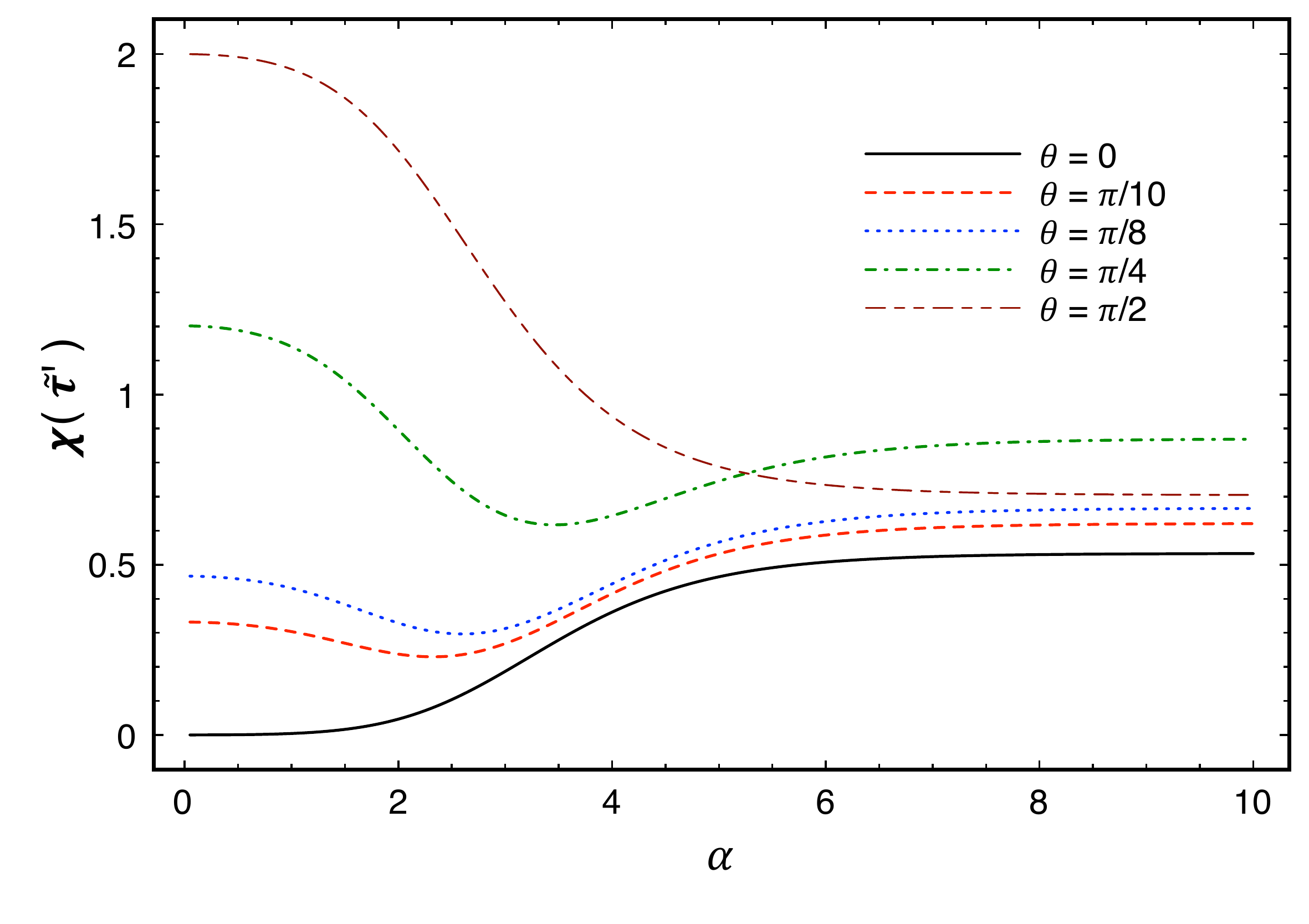}
\caption{\footnotesize{(Color online) The graph exhibits the Holevo bound $\chi(\tilde{\tau}')$ as a function of $\alpha=-\tanh^{-1}v$ for different values of the angle $\theta$. We have chosen $K_0=1$, $K_1=50$, $W_0=0.05$, $W_1=6$, and $\lambda_1=\lambda_2=\lambda_3=\lambda_4=1/4.$ For smaller angles ($\theta=0,$ $\theta=\pi/10$, and $\theta=\pi/8$), although $\chi(\tilde{\tau}')$ initially decreases with $\alpha$ it eventually increases as the rapidity gets larger, overcoming its value for $\alpha=0$ when $\alpha\rightarrow \infty$.}} 
\label{plot3}
\end{figure}
Now, using again the identity $S(\omega_1\otimes \omega_2)=S(\omega_1)+S(\omega_2)$, which is valid for any density matrices $\omega_1$ and $\omega_2$, we can write the Holevo bound
\begin{eqnarray}
\chi (\tilde{\tau}') &=& S (\tilde{\tau}') - \lambda_1 \, S (\tau_0'\otimes \tau_0')-\lambda_2 \, S (\tau_0'\otimes \tau_1') \nonumber \\
&-& \lambda_3 \, S (\tau_1'\otimes \tau_0')-\lambda_4 \, S (\tau_1'\otimes \tau_1')
\end{eqnarray} 
as 
\begin{eqnarray}
\chi (\tilde{\tau}') &=& -\sum_{\tilde{l}=1}^{4}\tilde{ \gamma}_{\tilde{l}} \, \log_{2} \tilde{\gamma}_{\tilde{l}} -2\lambda_1S(\tau_0')  - 2\lambda_4S(\tau_1') \nonumber \\
&-&(\lambda_2+\lambda_3)[S(\tau_0')+ S(\tau_1')], \\
\nonumber
\end{eqnarray}
where $\tilde{\gamma}_{\tilde{l}},$ $\tilde{l}=1,2,3,4,$ are the eigenvalues of $\tilde{\tau}'$. 

In Fig.~\ref{plot3} we plot  $\chi(\tilde{\tau}')$ as a function of the rapidity $\alpha$ for different values of the angle $\theta$. We have verified that the behavior of the Holevo bound for $\tilde{\tau}'$ as a function of $\alpha$ for different values of $K_1$ is very similar to that of $\chi(\tau')$, which is shown in Fig.~\ref{plot1}. Similarly, the behavior $\chi(\tilde{\tau}')$, when $\alpha \rightarrow \infty$, as a function of $W_1$ looks very close to the behavior of $\chi(\tau')$ described in Fig.~\ref{plot2}. Therefore, in order to analyze the Holevo bound for different choices of the angle $\theta$ of $\psi_1$, we have fixed $K_1$ and $W_1$ much larger than $K_0$ and $W_0$, respectively. Explicitly, we have chosen $K_0=1$, $K_1=50$, $W_0=0.05$, $W_1=6$, and $\lambda_1=\lambda_2=\lambda_3=\lambda_4=1/4.$ As can be seen from Fig.~\ref{plot3}, $\chi(\tilde{\tau}')$ initially decreases as $\alpha$ increases. However, for small angles (for instance, $\theta=0,$ $\theta=\pi/10$, and $\theta=\pi/8$), the Holevo bound begins to increase as $\alpha$ gets larger and, eventually, $\chi(\tilde{\tau}')$ overcomes its value for $\alpha=0$ (where Bob is at rest with respect to Alice). Thus, for some codifications of the classical bits, the (relativistic) velocity of the receiver can increase his accessible information. For angles closer to $\pi/2,$ Bob's movement only makes the Holevo bound smaller compared to the case where Alice and Bob do not have a relative motion. In particular, we can see from the graph that, when $\alpha$ is large enough, the use of orthogonal states, $\theta=\pi/2,$ is not even the best strategy anymore [i.e. orthogonal states do not maximize $\chi(\tilde{\tau}')$]. Thus, for larger angles, it is better to use both $W_0$ and $W_1$ much smaller than $1$ so that $\psi_0$ and $\psi_1$ are ``almost" momentum eigenstates and therefore, the effects of  the motion of the receiver on the accessible information are negligible.

It is interesting to note that, as in the case where two bits are being sent through the quantum channel, Alice can ``hide" the information on the momentum degrees of freedom by using states $\psi_{ij},$ $i,j=0,1$, with $\theta=0$ (which are completely indistinguishable in spin), and let Bob's movement make the information available to him in the spin degrees of freedom. As we have already pointed out in Section~\ref{twobitsrel}, this might be useful to protect the communication against possible noises in the spin degrees of freedom.

\section{Relativistic Quantum Map}
\label{quantummap}

\begin{figure}[t]
\begin{center}
\includegraphics[height=0.25\textheight]{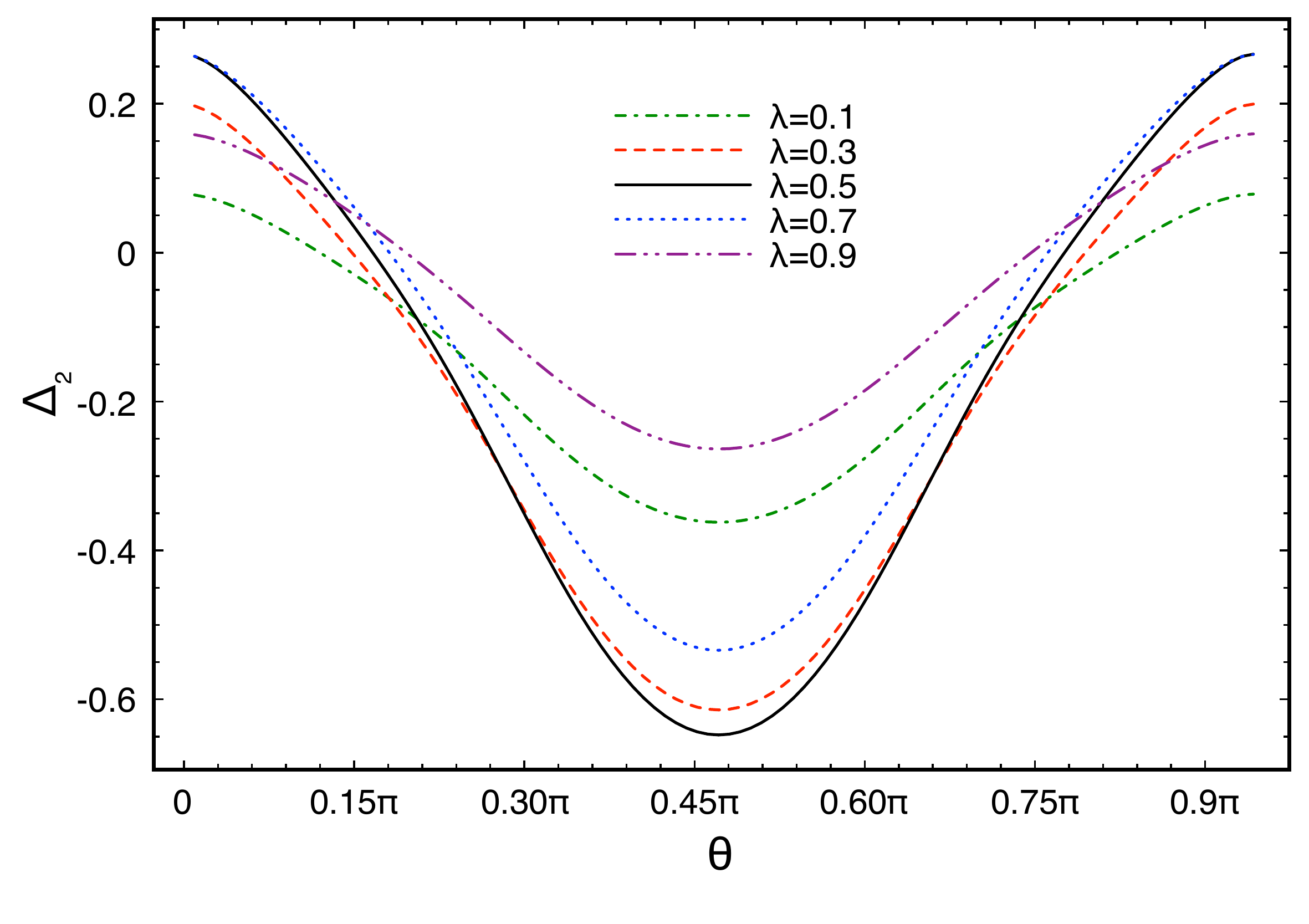}
\caption{(Color online) The graph shows $\Delta_{2} \equiv \chi\left[\mathcal{E}(\tau)\right]- \chi(\tau)$ as a function of $\theta$ for different values of $\lambda$ when $\alpha \rightarrow \infty$. We have fixed $W_0=0.05,$ $W_1=6,$ $K_0=1$, and $K_1=50$. We can see that there are two range of angles in which $\Delta_{2 }\geq 0$ and therefore the map $\mathcal{E}$ is not CP for angles within these ranges.}
\label{plot4}
\end{center}
\end{figure}
Let us  now define linear and trace-preserving maps that describe the relativistic quantum channels for both the transmission of two and four classical bits. For this purpose, we fix the values of $\theta \; (0<\theta<\pi),$ $K_i$, and $W_i$, $i=0,1,$ and define the convex sets    
\begin{equation}
\mathcal{U}\equiv \{\lambda \tau_0+ (1-\lambda)\tau_1| 0 \leq \lambda\leq 1\},
\label{stateset2}
\end{equation}
and
\begin{eqnarray}
\mathcal{W}\equiv \!\! && \!\!\!\bigl\{  \lambda_1  \tau_0\otimes \tau_0 + \lambda_2 \tau_0 \otimes \tau_1 + \lambda_3  \tau_1 \otimes \tau_0+ \lambda_4 \tau_1\otimes \tau_1 \nonumber \\  & | &  0 \leq \lambda_i \leq 1, \sum_{i=1}^4\lambda_i=1\bigr \},
\label{stateset4}
\end{eqnarray}
where $\tau_0$ and $\tau_1$ are given in Eqs.~(\ref{tau0}) and~(\ref{tau1}), respectively. 
If $\mathcal{B}(\mathcal{H})$ denotes the set of (bounded) operators over a Hilbert space $\mathcal{H}$, the maps $\mathcal{E}: \mathcal{U} \rightarrow \mathcal{B}(\mathbb{C}^2)$ and $\mathcal{N}: \mathcal{W} \rightarrow \mathcal{B}(\mathbb{C}^2\otimes \mathbb{C}^2)$ are defined as

\begin{figure}[t] 
\begin{center}
\includegraphics[height=0.25\textheight]{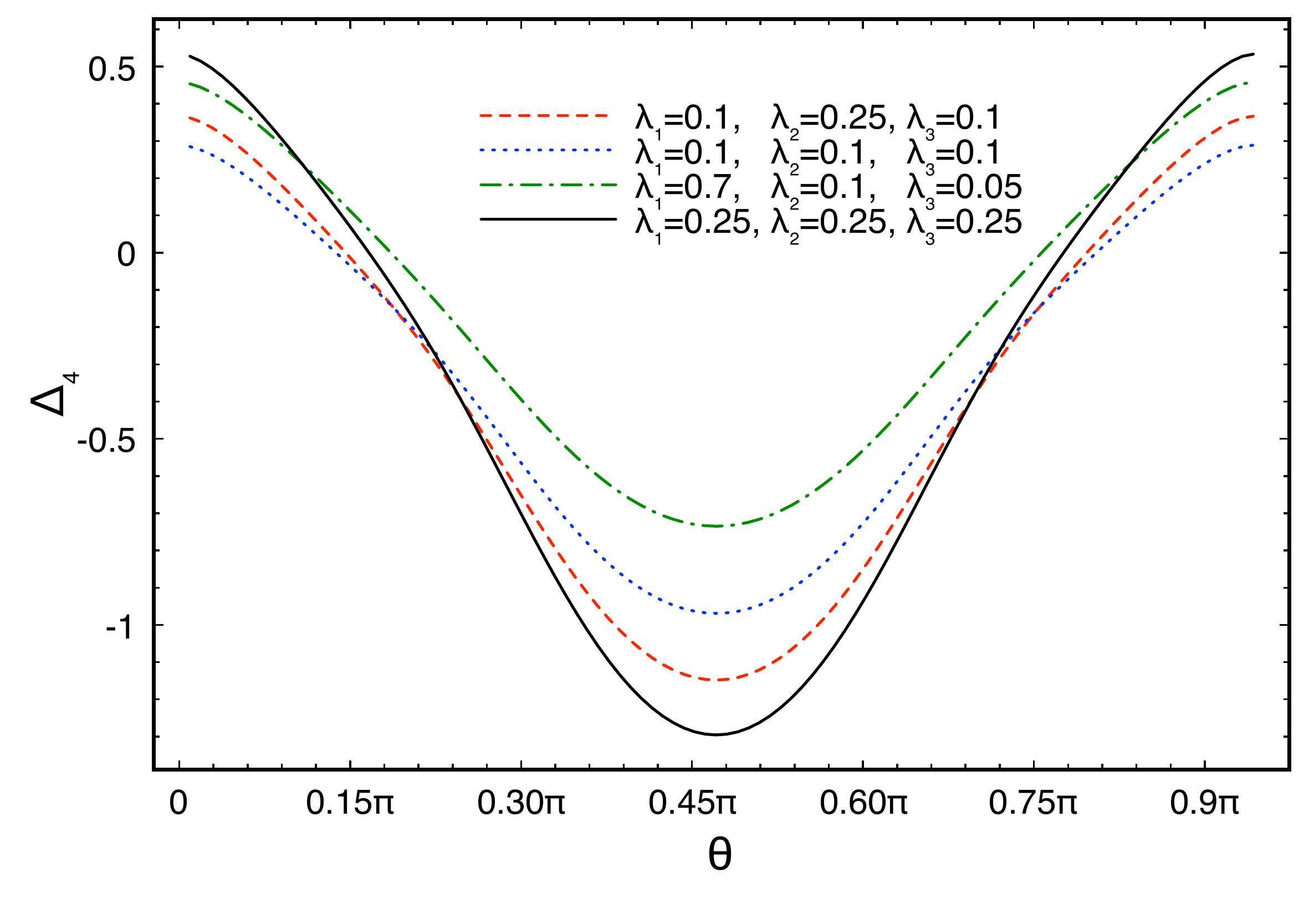}
\caption{(Color online) The graph shows $\Delta_{4} \equiv \chi\left[\mathcal{N}(\tilde{\tau})\right]- \chi(\tilde{\tau})$ as a function of $\theta$ for different values of $\lambda_1, \lambda_2, \lambda_3$ when $\alpha \rightarrow \infty$. We have fixed $W_0=0.05,$ $W_1=6,$ $K_0=1$, and $K_1=50$. We can see that there are two range of angles in which $ \Delta_{4} \geq 0$ and therefore the map $\mathcal{N}$ is not CP for angles within these ranges.}
\label{plot5}
\end{center}
\end{figure}

\begin{equation}
 \mathcal{E}(\tau)\equiv \int \mbox{d}{\bf p}\left[U(\Lambda)\rho({\bf p}, {\bf p})U^\dagger(\Lambda)\right]
\label{E}
\end{equation}
and 
\begin{equation}
 \!\mathcal{N}(\tilde{\tau})\! \equiv \!\!\! \int \!\!\!\mbox{d}{\bf p}\mbox{d}{\bf p}'\!\left[U(\Lambda)\!\otimes \!U(\Lambda) \tilde{\rho}({\bf p}, {\bf p}',{\bf p}, {\bf p}')U^\dagger(\Lambda)\!\otimes \! U^\dagger(\Lambda)\right]
\label{N}
\end{equation}
where $\rho \left( \mathbf{p}, \mathbf{\tilde{p}} \right)$, $\tilde{\rho} \left( \mathbf{p}, \mathbf{\tilde{p}},\mathbf{p}', \mathbf{\tilde{p}}' \right)$, and $U(\Lambda)$ are given in Eqs.~(\ref{completerho}),~(\ref{totdens4}), and~(\ref{booststate}), respectively. %~\footnote{Note that one can write the map $\mathcal{E}$ as $\mathcal{E}(\tau)={\rm tr}_{{\bf p}}\left[U(\Lambda)\mathcal{A}^{\theta}_{W_i,K_i}(\tau)U^{\dagger}(\Lambda)\right],$ for all $\tau\in \mathcal{U}.$ Here, $\mathcal{A}^{\theta}_{W_i,K_i}$ is a  convex-linear assignment map defined on $\mathcal{U}$ which satisfies $\mathcal{A}^{\theta}_{W_i,K_i}(\tau_i)({\bf p},{\bf p}')=\psi_i({\bf p})\psi^\dagger_i({\bf p}')\equiv \tau_i f_{{\bf k}_i}^{w_i}({\bf p})f_{{\bf k}_i}^{* w_i}({\bf p}')$.  It is easy to see that for the  four-bits case, the map $\mathcal{N}$ can be written as  $\mathcal{N}(\tilde{\tau})={\rm tr}_{{\bf p},{\bf \tilde{p}}}\left[U(\Lambda)\mathcal{G}^{\tilde{\tau}}_{W_i,K_i}(\tau)U^{\dagger}(\Lambda)\right],$ with $\mathcal{G}^{\theta}_{W_i,K_i}=\mathcal{A}^{\theta}_{W_i,K_i}\otimes\mathcal{A}^{\theta}_{W_i,K_i}$ for all $\tilde{\tau}\in \mathcal{W}$.}. Thus, by comparing Eqs.~(\ref{E}) and~(\ref{N}) with Eqs.~(\ref{tau`v2}) and~(\ref{tildetau'v2}), we conclude that $\mathcal{E}(\tau)=\tau'$ and $\mathcal{N}(\tilde{\tau})=\tilde{\tau}'$ for all $\tau\in \mathcal{U}$ and $\tilde{\tau} \in\mathcal{W}$. 

It is interesting to note that  when $W_0\ll W_1$ and $K_0\ll K_1$, the total state prepared by Alice, $\rho({\bf p}, {\bf \tilde{p}})$ for the two-bit case and $\rho({\bf p}, {\bf \tilde{p}},{\bf p}', {\bf \tilde{p}}')$ for the four-bit case, not only presents correlations between the spin and momentum degrees of freedom but also have non-vanishing quantum discord $\mathcal{D}$ (with respect to measurements made on the spin degrees of freedom). For the sake of simplicity, let us prove this statement for the state $\rho({\bf p}, {\bf \tilde{p}})$. However, a complete analogous calculation also shows the non-vanishing of the quantum discord in the four-bits case. We first note that

\begin{equation}
\left[\tau\otimes I_{\bf p}, \rho \right] = \lambda(1-\lambda)\cos \theta \left(\phi_{\uparrow}\phi_{\theta}^{\dagger}-\phi_{\theta}\phi_{\uparrow}^{\dagger} \right)\otimes(\rho_1-\rho_0),
\label{discord}
\end{equation}
where $I_{\bf p}$ is the identity in the momentum space, 
$$\tau\equiv \int \mbox{d}{\bf p}\rho({\bf p}, {\bf p}) \in \mathcal {U},$$ and, in the momentum representation, $\rho_i({\bf p}, {\bf p}')\equiv f_{{\bf k}_i}^{w_i}({\bf p})f_{{\bf k}_i}^{* w_i}({\bf p}'),$ i.e., they are the reduced momentum density operator associated with $\psi_i$. Now, we can see from Eq.~(\ref{discord}) that whenever $\rho_0\neq \rho_1$ (which is the case when $W_0\ll W_1$ and $K_0\ll K_1$) the above commutator does not vanish and thus, as shown in~\cite{facca10}, the state has non-vanishing quantum discord. Although it is not a sufficient condition, an initial state with non-vanishing discord indicates that the reduced spin dynamics ($\mathcal{E}$ and $\mathcal{N}$) may not be completely positive (CP) \cite{bdmrr12,mbcpv12, sl09}. 

Before we proceed to investigate this possibility, it is important to remark that one must be careful in defining an effective dynamics when there are some prior correlations between the system of interest and its environment, which in our case are the spin and momentum degrees of freedom of spin $1/2$ fermions, respectively. For example, even if the map  describing the effective dynamics is linear and trace preserving, it may fail to be positive. Thus, to define a physically reasonable dynamics, one must restrict the domain of the map to a set in which it takes positive operators into positive operators \cite{mbcpv12, jss04, ctz08, mra12, rmg10}.

The maps $\mathcal{E}$ and $\mathcal{N}$, given in Eqs.~(\ref{E}) and~(\ref{N}), describe the relativistic quantum channels for transmitting two and four bits of classical information, respectively. It is easy to see that they are convex-linear,  trace preserving (and can be extended, in a non-unique way, to linear and trace preserving maps acting on all linear operators), and positive. Thus, at least on $\mathcal{U}$ and $\mathcal{W}$, the maps $\mathcal{E}$ and $\mathcal{N}$ are well defined and describe the effective spin dynamics (which is enough for our purposes). It is well know that for any quantum map, i.e., any linear, trace-preserving, and CP operator $\mathcal{K}:\mathcal{B}(\mathcal{H})\rightarrow \mathcal{B}(\mathcal{H})$, where $\mathcal{H}$ is a Hilbert space, the Holevo bound satisfies \cite{nielsen&chuang}
\begin{equation}
\chi\left[\mathcal{K}(\rho)\right] \leq \chi(\rho),
\end{equation}
where $\rho$ is any density operator defined on $\mathcal{H}.$
Thus, quantum maps cannot increase the accessible information on the receiver. In Figs.~\ref{plot4} and~\ref{plot5} we plot 
\begin{equation}
\Delta_{2} \equiv \chi\left[\mathcal{E}(\tau)\right]- \chi(\tau)
\end{equation}
and 
\begin{equation}
\Delta_{4} \equiv \chi\left[\mathcal{N}(\tilde{\tau})\right]-\chi(\tilde{\tau})
\end{equation} 
as functions of $\theta$ for different probability distributions $\{\lambda, 1-\lambda\}$ and $\{\lambda_1, \lambda_2, \lambda_3, \lambda_4\}$, respectively. In both figures we have fixed $W_0=0.05,$ $W_1=6,$ $K_0=1,$ and $K_1=50$. We can see from the plots that there are cases in which 
\begin{equation}
\chi\left[\mathcal{E}(\tau)\right] \geq \chi(\tau)
\end{equation}
and 
\begin{equation}
\chi\left[\mathcal{N}(\tilde{\tau})\right] \geq \chi(\tilde{\tau}).
\end{equation}
Therefore, in the relativistic case, if one wants to always maximize the accessible information on the receiver, there will be quantum channels that cannot be described by quantum maps, i.e. linear, trace-preserving, and CP operators (the impossibility to describe quantum channels by CP maps in relativistic setups using photons as information carriers was noticed in \cite{PT02}). This will be the case when the classical information is encoded in states $\{\psi_0, \psi_1\}$ with the angle $\theta$ characterizing the state $\psi_1$ being smaller (larger) than some angle $\vartheta$ ($\pi - \vartheta$), with the value of $\vartheta$ depending on the number of bits and on their probability distribution, as can be seen from Figs.~\ref{plot4} and~\ref{plot5}.

For angles closer to $\pi/2$ however, as we have already pointed out,  the best strategy is to keep both $W_0$ and $W_1$ small to minimize the effects of the boost on the states. In the particular case where $W_0=W_1$, $K_0=K_1$, and $\theta=\pi/2$, it is easy to see that both $\mathcal{E}$ and $\mathcal{N}$ are $CP$ maps with Krauss decompositions 
\begin{equation}
\mathcal{E}(\tau) =\sum_{\mu=1}^3\Gamma_\mu \tau \Gamma_\mu
\end{equation}
\begin{equation}
\mathcal{N}(\tilde{\tau})=\left(\mathcal{E}\otimes\mathcal{E}\right)(\tilde{\tau})=\sum_{\mu,\nu=1}^3\left[\Gamma_\mu\otimes\Gamma_\nu\right] \tilde{\tau}\left[\Gamma_\mu\otimes\Gamma_\nu\right] 
\end{equation}
respectively, where the Krauss operators $\Gamma_\mu$ are given by
\begin{eqnarray}
\Gamma_1 \equiv \sqrt{1-V(\alpha)}\; I, \Gamma_2 \equiv \sqrt{\frac{V(\alpha)}{2}}\;\sigma_{\rm x}, \; \Gamma_3 \equiv \sqrt{\frac{V(\alpha)}{2}}\;\sigma_{\rm y}, \nonumber 
\end{eqnarray}
with $I$ being the identity operator and $V(\alpha)$ being given in Eq.~(\ref{Valpha}).

Even in the CP regime, when one is dealing with quantum process tomography the initial correlations between the system of interest and its environment must be treated carefully. In such cases the preparation procedure plays a crucial role and the tomographically reconstructed quantum map may differ from the dynamical quantum map \cite{kmrs07, bgtw11}.  It would be very interesting to investigate such issues in these relativistic scenarios. 
   
\section{Final Remarks}
\label{finalremarks}

In the present paper, we have used the Holevo bound to analyze how the relative motion between the sender and the receiver influences the capacity of a quantum communication channel to convey classical information. To this end, we have assumed that the sender, Alice, encodes the classical information in the spin degrees of freedom of spin-$1/2$ fermions of mass $m$ and sends the state prepared to the receiver, Bob, who is moving with velocity $v=-\tanh\alpha$ with respect to her. Bob then makes a spin measurement on the state and has to identify the message sent by Alice based on its measurement outcome. 

First it was analyzed the case where Alice has a classical information source that produces symbols $X=0,1$ according to the probability distribution $p_0, p_1$. Depending on the value of $X$, Alice prepares the spin-$1/2$ particle in a pure quantum state $\psi_X$ and sends it to Bob. The spin part of $\psi_0$ and $\psi_1$ were assumed to be eigenstates, with eigenvalue $1/2$, of $S_{\rm z}$ and ${\bf S}\cdot {\bf n}$, with ${\bf n}=(\sin 2\theta,0,\cos 2\theta)$,  respectively. It was shown that when $\theta$ is ``close" to $ \pi/2$, Bob's movement always reduces the Holevo bound and thus, the best strategy in this case is to use very narrow wave packets in the momentum degrees of freedom. This way, the states $\psi_0$ and $\psi_1$ are almost momentum eigenstates and therefore the effects of the boost on them are negligible. For small angles however, if Alice chooses $W_0 \ll W_1$ and $K_1$ large enough, and Bob moves sufficiently fast, the Holevo bound $\chi(\tau')$ overcomes $\chi(\tau)$,  the Holevo bound in the case where Bob is at rest relative to Alice. 

We have also analyzed how the above results generalize to the case where more classical bits are sent through the quantum channel. We have shown explicitly in the case where Alice has an information source  that produces four bits $\tilde{X}=00,01,10,11$ according to the probability distribution $p_{00},p_{01},p_{10},p_{11}$ and codifies each bit in one of the product states $\psi_i\otimes\psi_j,$ $i,j=0,1,$ that the conclusions reached in the two bit case can be extended to this one.

The use of non-orthogonal quantum states to convey classical information is important not only due to 
 cryptographic purposes but  also due to the fact that there are noisy quantum channels in which the optimal rate of information transmission is achieved only by using non-orthogonal states \cite{fucs97}. Our results seem to indicate that when the classical information is conveyed through such noisy quantum channels and the sender and receiver are in relativistic  relative motion, one might actually increase the channel capacity by carefully preparing the momentum degrees of freedom of the states. As we have shown, this is possible because, relativistically, boosts entangle the spin and momentum degrees of freedom of a spin-$1/2$ particle. Therefore, owing to his motion, the receiver can obtain some extra information in the spin degrees of freedom due to the fact that the states prepared by the sender are more distinguishable in momentum than in spin.

\begin{acknowledgments}
The authors are indebted to George Matsas for useful discussions and for reading the manuscript. We also thank Roberto Serra for his useful comments and for pointing out Ref. [40] to us. A.L. and A.T. acknowledge partial and full support from the Brazilian National Institute for Science and Technology of Quantum Information (INCT-IQ) and Coordena\c c\~ao de Aperfei\c coamento de Pessoal de N\'ivel Superior (CAPES) respectivelly. 
\end{acknowledgments}

\end{document}